
\input harvmac

\def\ack{\bigbreak\bigskip\centerline{{\bf Acknowledgements}}\nobreak
\bigskip}

\sfcode`\?=3000

%
%


\def\MPL#1#2#3{{Mod.~Phys.~Lett.} A#1 (19#2) #3}
\def\IJMP#1#2#3{{Int.~J.~Mod.~Phys.} A#1 (19#2) #3}
\def\PL#1#2#3{{Phys.~Lett.} B#1 (19#2) #3}
\def\NP#1#2#3{{Nucl.~Phys.} B#1 (19#2) #3}
\def\CMP#1#2#3{{Commun.~Math.~Phys.} #1 (19#2) #3}

\def\JETP#1#2#3{{Sov.~Phys.~JETP.} #1 (19#2) #3}
\def\ZETF#1#2#3{{Zh.~Eksp.~Teor.~Fiz.} #1 (19#2) #3}


\def\frac#1#2{{#1 \over #2}}
\def\inv#1{{1 \over #1}}
\def\ha{{1 \over 2}}

\def\ket#1{ \vert#1\rangle}
\def\->{\rightarrow}     \def\<-{\leftarrow}
\def\<{\langle}          \def\>{\rangle}
\def\[{\left [}          \def\]{\right ]}
\def\({\left (}          \def\){\right )}
\def\makeblankbox#1#2{\hbox{\lower\dp0\vbox{\hidehrule{#1}{#2}%
   \kern -#1
   \hbox to \wd0{\hidevrule{#1}{#2}%
      \raise\ht0\vbox to #1{}
      \lower\dp0\vtop to #1{}
      \hfil\hidevrule{#2}{#1}}%
   \kern-#1\hidehrule{#2}{#1}}}%
}%
\def\hidehrule#1#2{\kern-#1\hrule height#1 depth#2 \kern-#2}%
\def\hidevrule#1#2{\kern-#1{\dimen0=#1\advance\dimen0 by #2\vrule
    width\dimen0}\kern-#2}%
\def\openbox{\ht0=1.2mm \dp0=1.2mm \wd0=2.4mm  \raise 2.75pt
\makeblankbox {.25pt} {.25pt}  }

\def\bun#1/#2{\leavevmode
   \kern.1em \raise .5ex \hbox{\the\scriptfont0 #1}%
   \kern-.1em $/$%
   \kern-.15em \lower .25ex \hbox{\the\scriptfont0 #2}%
}
\def\row#1#2{#1_1,\ldots,#1_#2}
\def\apar{\noalign{\vskip 2mm}}

\def\opensquare{\ht0=3.4mm \dp0=3.4mm \wd0=6.8mm  \raise 2.7pt
\makeblankbox {.25pt} {.25pt}  }


\def\sector#1#2{\ {\scriptstyle #1}\hskip 1mm
\mathop{\opensquare}\limits_{\lower 1mm\hbox{$\scriptstyle#2$}}\hskip 1mm}

\def\tsector#1#2{\ {\scriptstyle #1}\hskip 1mm
\mathop{\opensquare}\limits_{\lower 1mm\hbox{$\scriptstyle#2$}}^\sim\hskip 1mm}


\def\op#1{\mathop{\rm #1}\nolimits}

\def\Det{\op{Det}}


\def\bZ{{\bf Z}}


\def\a{\alpha}
\def\b{\beta}

\def\d{\delta}          
\def\e{\epsilon}
\def\g{\gamma}          
\def\la{\lambda}                
\def\m{\mu}

\def\o{\omega}

\def\t{\theta}          

\def\vt{\vartheta}

%
              
              \def\CF{{\cal F}}
       \def\CH{{\cal H}}       
              
\def\CM{{\cal M}}              \def\CO{{\cal O}}
              \def\CR{{\cal R}}

%



\lref\rAKMW{O.~Alvarez, T.~Killingback, M.~Mangano and P.~Windey,
Commun. Math. Phys. 111 (1987) 1;
 Nucl.~Phys.~B (Proc.~Suppl.) 1A (1987) 189}
\lref\rCIZ{A.~Cappelli, C.~Itzykson, and J.-B.~Zuber,  Nucl.~Phys. B280 [FS 18]
(1987) 445; \CMP{113}{87}{1}\semi
A.~Kato, \MPL{2}{87}{585}}
\lref\rDY{ P.~Di Francesco and S.~Yankielowicz,
         ``Ramond Sector Characters and N=2 Landau-Ginzburg
   Models,'' Saclay preprint SPhT 93/049 and Tel-Aviv University preprint TAUP
2047-93 (hep-th.9305037)}
\lref\rEGH{T.~Eguchi, P.B~Gilkey and A.J~Hanson, Phys.~Rep. 66 (1980) 213}
\lref\rEOTY{T.~Eguchi, H.~Ooguri, A.~Taormina and S.-K.~Yang,
\NP{315}{89}{193}}
\lref\rET{T.~Eguchi and A.~Taormina, \PL{200}{88}{315}; \PL{210}{88}{125}}
\lref\rEY{T.~Eguchi and S.-K.~Yang, \MPL{5}{90}{1693}}
\lref\rEZ{M.~Eichler and D.~Zagier, ``The Theory of Jacobi Forms,''
(Birkh\"auser, 1985)}
\lref\rFGLS{P.~Fr{\' e}, F.~Gliozzi, M.R.~Monteiro and A.~Piras,
Class.~Quantum.~Grav. 8 (1991) 1455;
P.~Fr{\' e}, L.~Girardello, A.~Lerda and P.~Soriani,
        \NP{387}{92}{333}}
\lref\rGepner{D.~Gepner, \NP{296}{88}{757}}
\lref\rHirzebruch{F.~Hirzebruch, ``Elliptic genera of level $N$ for
complex manifolds,'' MPI preprint, MPI/88-24}
\lref\rLVW{W.~Lerche, C.~Vafa, and N.P.~Warner, \NP{324}{89}{427}}
\lref\rLW{W.~Lerche and N.P.~Warner, \PL{205}{88}{471}}
\lref\rMartinec{E.~Martinec, \PL{217}{89}{431}; ``Criticality, Catastrophes,
and
       Compactifications,''
      in {\it Physics and Mathematics of Strings}, ed. L.~Brink, D.~Friedan,
      and A.~M.~Polyakov (World Scientific, 1990)}
\lref\rMinimal{A.B.~Zamolodchikov and V.A.~Fateev, \JETP{63}{86}{913}
             \hfil\break
            [\ZETF{90}{86}{1553}]\semi
W.~Boucher, D.~Friedan and A.~Kent, \PL{172}{86}{316}\semi
P.~Di Vecchia, J.L.~Petersen, M.~Yu and H.B.~Zheng, \PL{174}{86}{280}\semi
F.~Ravanini and S.-K.~Yang, \PL{195}{87}{202}\semi
Z.~Qiu, \PL{188}{87}{207}; \PL{198}{87}{497}}
\lref\rMohri{K.~Mohri, private communication}
\lref\rO{S.~Odake, \MPL{4}{89}{557}; \IJMP{5}{90}{897}}
\lref\rSS{A.~Schwimmer and N.~Seiberg, \PL{184}{87}{191}}
\lref\rSW{A.~Schellekens and N.P.~Warner, \PL{177}{86}{317};
\NP{287}{87}{317}\semi
K.~Pilch, A.~Schellekens and N.P.~Warner, \NP{287}{87}{362}  }
\lref\rVafa{C.~Vafa, \MPL{4}{89}{1169}}
\lref\rVW{C.~Vafa and N.P~Warner, \PL{218}{89}{51}}
\lref\rWitten{E.~Witten, \NP{202}{82}{253}}
\lref\rWitteni{E.~Witten, \CMP{109}{87}{525}; ``The Index of the Dirac
Operator in Loop Space,'' in P.S.~Landweber, ed., {\it Elliptic Curves and
Modular Forms in Algebraic Topology} (Springer-Verlag, 1988)}
\lref\rWittenii{E.~Witten, ``On the Landau-Ginzburg description of $N=2$
 minimal models,'' IAS preprint IASSNS-HEP-93/10 (hep-th.9304026)}
\lref\rWitteniii{E.~Witten, ``Phases of $N=2$ theories in two dimensions,''
IAS preprint IASSNS-HEP-93/3 (hep-th.9301042)}



\noblackbox
\vbadness=10000
\Title{\vbox{\baselineskip12pt\hbox{KEK-TH-362}
                             \hbox{KEK preprint 93-51}
                             \hbox{hep-th/9306096}
                              \hbox{June 1993}}}
{
\vbox{\vskip -2.5cm\centerline{ Elliptic Genera }
\bigskip
\centerline{and}
\bigskip
\centerline{$N=2$ Superconformal Field Theory}}}
\centerline{%
Toshiya Kawai${}^\ast$, Yasuhiko Yamada${}^\ast\ {}^\$$ and
                 Sung-Kil Yang${}^\dagger$
\footnote{${}^\$ $}{Partially supported by Grant-in-Aid for Scientific Research
on Priority Area 231 ``Infinite Analysis''.}
}%
\it
\vskip.2in
\centerline{${}^\ast$National Laboratory for High Energy Physics (KEK)}
\centerline{Tsukuba, Ibaraki 305, Japan}
\medskip
\centerline{${}^\dagger$Institute of Physics, University of Tsukuba,}
\centerline{Ibaraki 305, Japan}
\rm
\vskip .4in
\centerline{\bf Abstract}
\medskip

Recently Witten proposed to consider elliptic genus in $N=2$ superconformal
field theory  to understand the relation between $N=2$ minimal models and
Landau-Ginzburg theories.   In this paper we  first discuss the basic
properties satisfied by elliptic genera  in  $N=2$ theories. These properties
are confirmed by some fundamental class of examples.
Then we introduce a generic procedure  to compute
the elliptic genera  of a particular class of orbifold theories, {\it i.e.\/}
 the ones orbifoldized  by $e^{2\pi iJ_0}$  in the Neveu-Schwarz sector.
This enables us to calculate the elliptic genera for Landau-Ginzburg orbifolds.
When the Landau-Ginzburg orbifolds allow an interpretation as target manifolds
with $SU(N)$ holonomy we can compare the expressions with the ones
obtained by  orbifoldizing tensor products of $N=2$ minimal models.
We also give  sigma model expressions of the elliptic genera for manifolds
of $SU(N)$ holonomy.
\Date{}
\footline={\hss\tenrm-- \folio\ --\hss}

\newsec{Introduction}

Over the past several years considerable number of studies have been made on
$N=2$ superconformal field theory (SCFT). Although the initial physical
motivation was to construct space-time supersymmetric vacua of string theory,
it has been gradually appreciated that what has been the source
of continual investigations is rooted in the theory's own profundity.
One notable manifestation  of this, and which is not unrelated to the
original motivation, is the connection of some $N=2$ SCFTs to the
Landau-Ginzburg models  or mathematically to singularity theory \rMartinec\rVW.
Quite recently, Witten \rWittenii\ cast a new light on this subject by
proposing to compare  elliptic genera in both theories\foot{
See also his previous work \rWitteniii.}. It is
somewhat curious that  elliptic genera in $N=2$ SCFT have
been paid  little attention
although the notion of elliptic genus was introduced long before
\rSW\rWitteni\rAKMW.

In  the present work  we first spell out what ought to be satisfied by
the elliptic genus of $N=2$ SCFT.
As $N=2$ superconformal algebra  contains the $U(1)$ current,
the elliptic genus of $N=2$ SCFT depends on  the $U(1)$ angle  besides
the modular parameter.  The $U(1)$ decoupling argument implies that the
elliptic genus must satisfy certain two-variable functional equations which
play a  crucial role throughout this paper.

It is well-known that in order to construct a viable string vacuum or
more generally in order for a $N=2$ SCFT to have some sigma model
interpretation we have to impose
$U(1)$ charge integrality \rGepner\ besides the condition that
one-third of the Virasoro central charge be an integer. It is also known that
this is achieved via orbifoldization
by $e^{2\pi iJ_0}$ in the Neveu-Schwarz sector (see for instance \rVafa).
We will establish a method to compute the elliptic genus for such a class of
 orbifolds. This, in particular,  enables us to compute the elliptic genera for
Landau-Ginzburg orbifolds \rVafa\ and to compare them with the results obtained
by the methods in \rEOTY\ in the case where manifold interpretation is present.
We also give a  sigma model expression of the elliptic genus
directly related to the geometry of the target manifold.

The organization of the paper is as follows.
In Sect.2 we summarize the fundamental properties of the elliptic genus of
$N=2$ SCFT.
In Sect.3 we give the elliptic genera of the $N=2$ minimal models and
Landau-Ginzburg models.  It is seen that they enjoy the properties derived in
Sect.2. Sect.4 explains how  to compute the elliptic genus of
the orbifold theory mentioned above. We then apply this formalism to
the examples presented in Sect.3. As a by-product, we can reproduce Vafa's
formula \rVafa\ for the Euler characteristic of  Landau-Ginzburg orbifolds.
Finally, in Sect.5, we investigate the properties and examples of
elliptic genera for theories corresponding to manifolds of $SU(N)$ holonomy
together with their sigma model expressions.

Appendix A and B summarize some useful formulae to be used in the text.

\newsec{Elliptic genus}

We consider the Ramond sector of an $N=2$ superconformal theory with central
charge $c$. Let $L_0$ ($\bar L_0$) be a Virasoro generator of
left (right)-movers
and $J_0$ ($\bar J_0$) be a $U(1)$ charge operator of left (right)-movers.
Following Witten \rWittenii\ we define the elliptic genus
\eqn\genus{
Z(\tau, z)=
\op{Tr} (-1)^F q^{L_0-{c/24}} ~  \bar q^{\bar L_0-{c/24}} ~
  y^{J_0}\, ,}
where $q=e^{2\pi i \tau}$ with $\tau$ defined on the upper half-plane,
$y=e^{2\pi i z}$ and
$F=F_L -F_R$ with $F_L$ ($F_R$) being the fermion number of
left (right)-movers. Here the trace is taken over the states in the Ramond
sector and the fermion parity operator in $N=2$ theories is given by
\eqn\?{
(-1)^F = \exp [i \pi (J_0 -\bar J_0)]\,,}
where we have assumed as usual that the difference of the left and right
$U(1)$ charges is an integer \rLVW. In \genus\ though we have written
$\bar q^{\bar L_0-{c/24}}$ explicitly to remind that we are
dealing with the combined left and right-moving sectors the elliptic
genus is of course independent of $\bar \tau$ by virtue of supersymmetry
of the Ramond sector. Setting $z=0$ in \genus\ we have the Witten
index
\eqn\index{
Z(\tau, 0)= \op{Tr}(-1)^F\,,}
which may yield the Euler characteristic of the target manifold
if the theory admits a sigma model interpretation \rWitten.

Now the fundamental properties of the elliptic genus are summarized.
First of all the spectrum of the Ramond sector is symmetric under
charge conjugation. Hence
\eqn\chargesym{
Z(\tau, z)=Z(\tau, -z)\,.}
We next discuss the modular property of the elliptic genus. Since the trace
is taken over the Ramond sector with  $(-1)^F$ insertion the boundary
condition for the elliptic genus is invariant under the $SL(2,{\bf Z})$
transformations.  Thus under $\tau \rightarrow \tau +1$ it follows
that
\eqn\?{
Z(\tau +1, z)=Z(\tau, z)\,,}
whereas under $\tau \rightarrow -1/\tau$
\eqn\?{
Z (-1/\tau, z /\tau)=
e^{2\pi i (\hat c/ 2)(z^2 /\tau)} Z(\tau, z)\,,}
where we have set $\hat c =c/3$ and the $z$-dependence is fixed by
the standard $U(1)$ decoupling argument in $N=2$ theories. Hence we find
\eqn\modular{
Z \( {a\tau +b \over c\tau +d}, {z \over c\tau +d} \)=
e^{2\pi i (\hat c /2) c z^2 /( c\tau +d)} Z(\tau, z)\,,\quad
\pmatrix{a&b\cr c&d\cr}\in SL(2,\bZ)\,.}
Note that the choice $a=d=-1$, $b=c=0$ gives charge conjugation symmetry
\chargesym.

Suppose that the $U(1)$ charges of the chiral ring elements in the
Neveu-Schwarz sector are multiples of $1/h$.
Then $\hat c h$ is an integer since
the top chiral ring element has the $U(1)$ charge $\hat c$.
We now wish to examine the periodicity with respect to the variable $z$.
For this  we invoke
the spectral flow \rSS\ which is the inner automorphism of the $N=2$ algebra
\eqn\specflow{\eqalign{
&L_n \rightarrow L_n+\lambda J_n
            + {\hat c \over 2} \lambda^2 \delta_{n,0} ~,  \cr
&J_n \rightarrow J_n +\hat c \lambda \delta_{n,0} ~,  \cr
&G_r^\pm \rightarrow G_{r \pm \lambda}^\pm ~, \cr}}
where $\lambda$ is an arbitrary parameter.
Since the $U(1)$
charge is shifted by $\hat c /2$  under the spectral flow
to the Ramond sector we find for $\mu \in h{\bf Z}$
\eqn\periodmu{
Z(\tau, z+\mu)=(-1)^{\hat c \mu}Z(\tau, z)\,.}
Furthermore under the spectral flow \specflow\ with
$\lambda \in h{\bf Z}$ the Ramond ground-states are mapped onto
themselves. This implies
\eqn\?{
Z(\tau, z+\lambda \tau)=(-1)^{\hat c \lambda}
 e^{-2\pi i (\hat c/2) (\lambda^2 \tau +2\lambda z)}Z(\tau, z)\,.}
Thus we obtain the double quasi-periodicity of the elliptic genus
\foot{ We note that
the transformation equations
\eqnn\period
\modular\ and \period\
correspond to some of the defining properties of a Jacobi form of weight 0 and
index $\hat c/2$ when $\hat c/2 \in {\bf Z}$ and $h=1$ \rEZ.
See also footnote below (5.19). }
$$
Z(\tau, z+\lambda \tau +\mu)=(-1)^{\hat c (\lambda +\mu)}
 e^{-2\pi i (\hat c/ 2) (\lambda^2 \tau +2\lambda z) }
Z(\tau, z)\,,\quad \lambda, \mu \in h{\bf Z}\,.\eqno\period
$$

To summarize, the properties \modular\ and \period\ together
with the limiting behavior as $\tau\->i\infty$ characterize the elliptic genus
({\it c.f.\/} \rDY).
In unitary theories the existence of  $\lim_{\tau\->i\infty}Z(\tau,z)$
is guaranteed.

\newsec{Examples of elliptic genus}

In this section we give some explicit examples of elliptic genera and
see that they indeed satisfy the fundamental properties mentioned in Sect.2.

\subsec{$N=2$ minimal models}

The $N=2$ minimal models with $\hat c=\frac{k}{k+2}$ $(k=1,2,...)$
have been extensively studied in the literature
\rMinimal.  The elliptic genera of diagonal ($A$-type) theories were
already obtained by Witten \rWittenii.
To construct elliptic genera for the $N=2$ minimal models in general
we consider the (twisted) characters of the Ramond representations
\eqn\?{
I_m^l(\tau,z)=\Tr_{\CH^R_{l,m}}(-1)^{F_L}q^{L_0-3\hat c/24}y^{J_0}=
(\chi_m^{l,1}-\chi_m^{l,-1})(\tau,z)\,,}
where $\chi^{l,s}_m$
($l\in\{0,\ldots,k\}$, $s\in\{-1,0,1,2\}$, $m\in\{-k-1,\ldots,k+2\}$)
are obtained through the  branching relation \rGepner
\eqna\?{
$$\eqalignno{
&\chi^l(\tau,w)\t_{s,2}(\tau,w-z)=\sum_{m=-k-1}^{k+2}
\chi^{l,s}_m(\tau,z)\t_{m,k+2}\left(\tau,w-2z/(k+2)\right)\,, &\?a\cr
&\chi^l(\tau,w)=\frac{\t_{l+1,k+2}-\t_{-l-1,k+2}}{\t_{1,2}-\t_{-1,2}}(\tau,w)
     =:\sum_{m=-k+1}^k c^l_m(\tau)\t_{m,k}(\tau,w)\, .&\?b\cr}$$}
For definitions and properties  of theta function see Appendix A.
The explicit formula of $\chi^{l,s}_m$ can be found  by use of (A.4)  as
\eqn\?{
\chi^{l,s}_m(\tau,z)=\sum\limits_{j\in {\bf Z}}
c^l_{m-s+4j}(\tau)
q^{(k+2)/(2k)\[m/(k+2)-s/2+2j\]^2}
y^{m/(k+2)-s/2+2j}\,.
}
It is well-known that the $N=2$ minimal model is constructed for each
simply-laced Lie algebra of Coxeter number $h=k+2$ through  a choice of
the Cappelli-Itzykson-Zuber matrix \rCIZ\ whose entries $N^{(k)}_{l\bar l}$ are
non-negative integers  satisfying
\eqn\propofN{\eqalign{
&\sum_{l,\bar l=0}^k N^{(k)}_{l\bar l}A_{ll'}^{(k)}A_{\bar l\bar l'}^{(k)}=
N^{(k)}_{l'\bar l'}\,,\cr
\apar
&N^{(k)}_{l\bar l}=0\,,
\quad\hbox{if $\frac{l(l+2)}{4(k+2)}\not\equiv
\frac{\bar l(\bar l+2)}{4(k+2)}\ \pmod{\bZ}$}\,,\cr
\apar
&N^{(k)}_{k-l,k-\bar l}=N^{(k)}_{l\bar l}\,,\cr}}
where
\eqn\?{
A_{ll'}^{(k)}=\sqrt{\frac{2}{k+2}}\sin\frac{\pi(l+1)(l'+1)}{k+2}\, .}
The elliptic genus for the $N=2$ minimal model is then given by
\eqn\MMgenus{
\eqalign{
Z(\tau,z)&=\ha\sum_{l,\bar l=0}^k\sum_{m=-k-1}^{k+2}N^{(k)}_{l\bar l}
I^l_m(\tau,z)
I^{\bar l}_m(\bar\tau,0)\cr
\apar
&=\sum_{l,\bar l=0}^k N^{(k)}_{l\bar l}I^l_{\bar l+1}(\tau,z)\,,\cr}}
{\noindent where} we used $I^l_m(\tau,0)=\d_{m,l+1}-\d_{m,-l-1}$ in the
second equality.
It is a straightforward calculation using the properties summarized in
Appendix B to check \modular\ and \period\ with $h=k+2$.

\subsec{Landau-Ginzburg models}

Consider the  $N=2$ Landau-Ginzburg  model \rMartinec \rVW\ with an action
\eqn\?{
\int d^2 z d^2 \theta d^2 {\overline \theta} X {\overline X}+
 (\int d^2 z  d^2 \theta W(X)+c.c.)\,,}
where the superpotential $W$ is a weighted homogeneous polynomial of
$N$ chiral superfields $X_1, \cdots, X_N$ with weights
$\omega_1, \cdots, \omega_N$,
\eqn\?{
\lambda W(X_1,\cdots,X_N)=
W(\lambda^{\omega_1} X_1,\cdots,\lambda^{\omega_N} X_N)\,.}
We assume that $W$ has an isolated critical point at the origin
and $\o_i$'s are strictly positive rational numbers such that $\row\o N\le\ha$.
Following Witten \rWittenii, the elliptic genus of the Landau-Ginzburg model
can be computed as
\eqn\LGgenus{
Z(\tau,z)=\prod_{i=1}^N Z_{\omega_i}(\tau,z)\,,}
where
\eqn\LGgenuspr{
\eqalign{
Z_{\omega}(\tau,z)&=y^{-(1-2\omega)/2} {(1-y^{1-\omega}) \over (1-y^{\omega})}
       \prod_{n=1}^{\infty}{(1-q^ny^{1-\omega}) \over (1-q^ny^{\omega})}
{(1-q^n y^{-(1-\omega)}) \over (1-q^n y^{-\omega})}  \cr
\apar
       &={\vartheta_1(\tau,(1-\omega) z) \over \vartheta_1(\tau,\omega z)}\,,
\cr}}
and $\vartheta_1(\tau,z)$ is one of the Jacobi theta functions.
Using the theta function formulae (see Appendix A)
it is easy to check the expected general properties \modular\ and \period\
of $Z(\tau, z)$
with $h$ being the smallest positive integer such that $\o_ih\in\bZ$ for
all $1\le i\le N$.
Note that the zero-mode part of the genus $Z(i \infty, z)$ is the
generating function of the  $U(1)$ charges of the Ramond vacua.

In \rWittenii, a remarkable connection between \LGgenus\
and some Ramond characters for  the $N=2$ algebra was pointed out.
This relation was  further elucidated in \rWittenii\ by constructing a
free field realization of the $N=2$ superconformal algebra. Witten's
construction using $N=2$ superfields  actually coincides with the
realization in terms of $\beta$-$\gamma$-$b$-$c$
earlier considered in \rFGLS.
Let us introduce free fields $\beta_i, \gamma_i, b_i, c_i$
of conformal weights $(1,0,1,0)$ with the OPEs given by
$\beta_i(z)\gamma_j(w)\sim \delta_{ij}/(z-w)$
 and $b_i(z)c_j(w)\sim \delta_{ij}/(z-w)$. Here we consider $\b_i,\g_i$
$(b_i,c_i)$ to be bosonic (fermionic).
For each $W$  one can associate a free field realization \rFGLS\ of the
twisted $N=2$ algebra
\foot{ Although we describe the
realization in twisted form for convenience, it is of course possible to
obtain the standard $N=2$ algebra by untwisting. }
\eqn\?{
\eqalign{
&J=\sum_{i=1}^N -(1-\omega_i) b_i c_i+\omega_i \beta_i \gamma_i\,,\cr
&G^+=\sum_{i=1}^N 2(1-\omega_i) \partial \gamma_i c_i
                  -2\omega_i \gamma_i \partial c_i\,,\cr
&G^-=\sum_{i=1}^N \beta_i b_i\,,\cr
&T=\sum_{i=1}^N \beta_i \partial \gamma_i-b_i \partial c_i\,.\cr}}
These satisfy the twisted (topological) $N=2$ algebra \rEY\
with $\hat c={c \over 3}=\sum_{i=1}^N (1-2 \omega_i)$.
The screening charge $Q={\overline Q_{+,L}}$
is defined by
\eqn\?{
Q=\oint dz \sum_{i=1}^N {\partial W(\gamma) \over \partial \gamma_i} b_i(z)\,,}
which is a holomorphic piece of one of the supercharges
${\overline Q_{+}}={\overline Q_{+,R}}+{\overline Q_{+,L}}$ of the
Landau-Ginzburg model.
It is straightforward to check that $Q$ commutes with the $N=2$ currents
\foot{Recently, some higher spin currents that commute with $Q$
are also constructed  in the classical limit \rMohri.}.
The genus \LGgenus\  can be identified with the index of $Q$
on the Fock space
$\CH=\otimes_{i=1}^N \CF(\beta_i, \gamma_i, b_i, c_i)$
\eqn\?{
\op{index}(Q)=\op{Tr}_{\CH}[(-1)^F q^{L_0} y^{J_0-\hat c /2}]\,,}
where the fermion number $F$ is $(0,0,1,-1)$ for $\beta_i, \gamma_i, b_i, c_i$.
Since the $N=2$ algebra acts on the $Q$-cohomology space,
the elliptic genus \LGgenus\ should be expressed as a  linear combination of
the $N=2$ characters.

It is easy to see that the chiral primary states $\CO\ket{\hbox{vac}}$,
$\CO\in{\bf C}[\g]/dW$ are nontrivial $Q$-cohomology elements.
If $W$ is of $A$-type, the whole $Q$-cohomology space is generated by such
chiral primaries over the $N=2$ algebra, however, in general case,
there are more generators than chiral primaries.
For instance, in the $D_4$ case we have
$Q=\oint dz [(\gamma_1^2+\gamma_2^2) b_1+2 \gamma_1 \gamma_2 b_2]$ and
the $Q$-cohomology space is generated by the six generators $1$, $\g_1$,
$\gamma_2$, $\gamma_1^2$, $\gamma_1 \beta_2+\gamma_2 \beta_1+2 (c_1 b_2+c_2
b_1)$ and $\g_1b_2-\g_2b_1$  over the $N=2$ algebra.
Each generator corresponds to the highest weight vector (annihilated by
$J_1,G^{-}_0$ and $G^{+}_1$) in the irreducible decomposition of genus
\MMgenus,
\eqn\?{
Z_{D_4}(\tau,z)=I^0_1+I^2_3+I^2_3+I^4_5+I^4_1+I^0_5\,.}

As conjectured by Witten \rWittenii,
if the superpotential $W$ is of  $ADE$ type,  the elliptic genus must
coincide with that of the corresponding  $N=2$  minimal model computed
above \MMgenus.
This is so since (adopting the reasoning in \rDY)  both  the expressions of
the elliptic genus satisfy the same \modular\ and \period\ and, as can be seen
easily, possess the identical limits as $\tau\->i\infty$.
In the simplest non-trivial case $k=1$ we can easily confirm this equality by
an explicit calculation.
Since   we  have
$I^1_2(\tau,z)=f_{\ha,3}(\tau,z/3)$
and $I^0_1(\tau,z)=f_{\ha,3}(\tau,-z/3)$, the elliptic genus \MMgenus\ is given
by $Z(\tau,z)=f_{\ha,3}(\tau,z/3)+f_{\ha,3}(\tau,-z/3)$. Here we  have
introduced
\foot{The meaning of the notation $f_{{1 \over 2},3}$ will become clear later.}
\eqn\defoff{
f_{{1 \over 2},3}(\tau, z)= {1 \over \eta (\tau )}
 \sum_{n\in \bZ} (-1)^n q^{(3/2)( n+1/6)^2}
 y^{3(n+1/6)}\, ,}
where the Dedekind $\eta$  function is given by
$\eta(q)=q^{1/24}\prod_{n=1}^\infty(1-q^n)$ and $y=e^{2\pi i z}$.
On the other hand, as a result of the classical quintuple product identity,
we obtain
\eqn\?{\eqalign{
f_{\ha,3}(\tau,z)&+f_{\ha,3}(\tau,-z)\cr
\apar
&=y^{-\ha}\prod_{n=1}^\infty(1+q^{n-1}y)
(1+q^ny^{-1})(1-q^{2n-1}y^2)(1-q^{2n-1}y^{-2})\cr
\apar
&=y^{-\ha}\prod_{n=1}^\infty\frac{(1-q^{n-1}y^{2})
(1-q^{n}y^{-2})}{(1-q^{n-1}y)(1-q^{n}y^{-1})}\, ,\cr}}
thus proving Witten's conjecture in this particular case.

\newsec{Orbifoldized elliptic genus}

Now that we have deduced the general properties of the elliptic genus of
$N=2$ SCFT and have seen some fundamental class of examples,
 in this section we wish to introduce a generic procedure, starting with
the elliptic genera of untwisted theories, to compute
the elliptic genera  of a particular class of orbifold theories, {\it i.e.\/}
 the ones orbifoldized  by $e^{2\pi iJ_0}$ in the Neveu-Schwarz sector.
This type  of orbifolds have been intensively studied by
string theorists since for $\hat c=$ integer it is
equivalent to Gepner's method \rGepner\ of
 constructing space-time supersymmetric string vacua.

\subsec{Procedure to compute orbifoldized elliptic genus}

Suppose we are given an arbitrary $N=2$ SCFT whose elliptic genus $Z(\tau,z)$
satisfies \modular\ and \period.
The contribution of the untwisted sector to the orbifoldized elliptic genus
is apparently given by
\eqn\?{
\sector00(\tau,z)=Z(\tau,z)\,.}
To describe the contribution of the twisted $\a$-sector projected by $\b$
we introduce
\eqn\absector{
\  \sector{\b}{\a} (\tau,z)=
e^{2\pi i(\hat c/2)\a\b}e^{2\pi i(\hat c/2)(\a^2\tau+2\a z)}
Z(\tau,z+\a\tau+\b)\,,\quad \a,\b\in\bZ\, .}
Definition  \absector\ is  motivated by the fact that twisted sectors
should  be obtained by integral amount of spectral flows.
It follows immediately from \absector\ that
\eqna\propofsector{
$$\eqalignno{
\apar
\sector\b\a(-1/\tau,z/\tau)&=
e^{2\pi i(\hat c/2)(z^2/\tau)}\sector{-\a}\b(\tau,z)\,,
   &\propofsector a\cr
\noalign{\vskip 3mm}
\sector\b\a(\tau+1,z)&=\sector{\a+\b}\a(\tau,z)\,,&\propofsector b\cr
\noalign{\vskip 3mm}
\sector\b\a(\tau,z+\la\tau+\m)&=e^{2\pi i(\hat c/2)(\a\m-\b\la-\la\m)}
                               e^{-2\pi i(\hat c/2)(\la^2\tau+2\la z)}&\cr
&\hskip 1.5cm
\times\sector{\b+\m}{\a+\la}(\tau,z)\,,\quad \la,\m\in \bZ\, .
&\propofsector c\cr}$$}
We now set
\eqn\?{
\tsector\b\a(\tau,z)=\e(\a,\b)\sector\b\a(\tau,z)\,,}
and determine $\e(\a,\b)$ by imposing
\eqna\?{
$$\eqalignno{
\tsector\b\a(-1/\tau,z/\tau)&=
e^{2\pi i(\hat c/2)(z^2/\tau)}\tsector{-\a}\b(\tau,z)\,,
   &\?a\cr
\noalign{\vskip 3mm}
\tsector\b\a(\tau+1,z)&=\tsector{\a+\b}\a(\tau,z)\,,&\?b\cr
\noalign{\vskip 3mm}
\tsector{\b+\m}{\a+\la}(\tau,z)&=\tsector\b\a(\tau,z)\,,\quad
\la,\m\in h\bZ\,, &\?c\cr}$$}
\vskip 0.1mm
\noindent{together} with the condition $\e(0,0)=1$.
A possible solution we will choose in the following is
\eqn\solofepsilon{
\e(\a,\b)=(-1)^{D(\a+\b+\a\b)}\,,}
where $D$ is some integer satisfying
\eqn\?{
Dh\equiv\hat c h \pmod{2}\,.}
It is obvious that the orbifoldized elliptic genus defined by
\eqn\orbifold{
Z_{\hbox{orb}}(\tau,z)=\inv{h}\sum_{\a,\b=0}^{h-1}\tsector\b\a(\tau,z)\,,}
obeys the same modular transformation laws as $Z$:
\eqn\?{
Z_{\hbox{orb}}\(\frac{a\tau+b}{c\tau+d},\frac{cz}{c\tau+d}\)=e^{2\pi i(\hat
c/2)
c z^2/(c\tau+d)}Z_{\hbox{orb}}(\tau,z)\,,\quad
\pmatrix{a&b\cr c&d\cr}\in SL(2,\bZ)\,.}
Furthermore, as a consequence of  \propofsector{c}, $Z_{\hbox{orb}}$ enjoys
the same double quasi-periodicity as $Z$:
\eqn\?{
Z_{\hbox{orb}}(\tau,z+\la\tau+\m)=(-1)^{\hat c (\la+\m)}
e^{-2\pi i(\hat c/2)(\la^2\tau+2\la z)}Z_{\hbox{orb}}(\tau,z)\,,
\quad \la,\m\in h\bZ\,.}

If $\hat c$ is an integer we can take $D=\hat c$ and in this
case we have a stronger quasi-periodicity
\eqn\?{
Z_{\hbox{orb}}(\tau,z+\la\tau+\m)=(-1)^{\hat c (\la+\m)}
e^{-2\pi i(\hat c/2)(\la^2\tau+2\la z)}Z_{\hbox{orb}}(\tau,z)\,,}
for {\it any\/} integers $\la$ and $\m$.
This shows that  after orbifoldization  only integral
(half-odd integral) charges survive in the Ramond sector
if $\hat c$ is even (odd).
Later we will investigate the  case $\hat c =$integer in more detail.

\subsec{Orbifolds of $N=2$ minimal models and their tensor products}

The elliptic genus of the $\bZ_{h=k+2}$ orbifold of the $N=2$ minimal model
is given by
\eqn\?{
Z_{\hbox{orb}}(\tau,z)=
\inv{h}\sum_{\a,\b=0}^{k+1}\ha\sum_{l,\bar l=0}^k\sum_{m=-k-1}^{k+2}
\xi^{(m+\a)\b}N^{(k)}_{l\bar l} I^l_m(\tau,z)I^{\bar l}_{m+2\a}(\bar\tau,0)\,.}
Using the properties of $I_m^l(\tau,z)$ it is easy to see that
$Z_{\hbox{orb}}(\tau,z)=-Z(\tau,z)$ which is the well-known self-duality of
the $N=2$ minimal model. It is also not difficult to rewrite this expression
 using the formulae in Appendix B as
\eqn\?{
Z_{\hbox{orb}}(\tau,z)=\inv{h}\sum_{\a,\b =0}^{k+1}(-1)^{\a+\b+\a\b}
\sector\b\a(\tau,z)\,,
}
which is in agreement  with  \orbifold\ since we can choose $D=1$.

The elliptic genus of tensor products of the $N=2$ minimal model is obviously
given by
\eqn\?{
Z(\tau,z)=\prod_{i=1}^r\sum_{l_i,\bar l_i=0}^{k_i} N^{(k_i)}_{l_i\bar l_i}
I^{l_i}_{\bar l_i+1}(\tau,z)\,. }
Then the elliptic genus of its orbifold turns out to be
\eqn\?{
Z_{\hbox{orb}}(\tau,z)=
\inv{h}\sum_{\a,\b=0}^{h-1}\prod_{i=1}^r
\ha\sum_{l_i,\bar l_i=0}^{k_i}\sum_{m_i=-k_i-1}^{k_i+2}
\xi_i^{(m_i+\a)\b}N^{(k_i)}_{l_i\bar l_i} I^{l_i}_{m_i}
(\tau,z)I^{\bar l_i}_{m_i+2\a}(\bar\tau,0)\,,}
where $\xi_i=e^{2\pi i/(k_i+2)}$ and
$h=\mathop{\hbox{LCM}}(k_1+2,\ldots,k_r+2)$.
This expression can also be rewritten in the form \orbifold\ with $D=r$.

\subsec{Landau-Ginzburg orbifolds}

One obtains  the elliptic genera of the Landau-Ginzburg orbifolds \rVafa\
by simply substituting \LGgenus\ into \orbifold.
Explicitly we have
\eqn\?{
Z_{\hbox{orb}}(\tau,z)=\inv{h}\sum_{\a,\b=0}^{h-1}
\e(\a,\b)e^{2\pi i(\hat c/2)\a\b}e^{2\pi i(\hat c/2)(\a^2\tau+2\a z)}
\prod_{i=1}^NZ_{\o_i}(\tau,z+\a\tau+\b)\,.}
As a non-trivial check of our formalism  we now evaluate the Witten index of
the Landau-Ginzburg orbifolds.
For this purpose it suffices to know
\eqna\?{
$$\eqalignno{
\apar
&\frac{\vt_1(\tau,(1-\o_i)(\a\tau+\b))}{\vt_1(\tau,\o_i(\a\tau+\b))}
=&\cr
\apar
&\hskip 1.7cm (-1)^{\a+\b}(-1)e^{2\pi i\o_i\a\b}
e^{-2\pi i\ha(1-2\o_i)\a^2\tau}\,, \quad
\hbox{if $i\not\in S_{\a\b}$}\,,&\?a\cr
\noalign{\vskip 3mm}
&\lim_{z\->0}
\frac{\vt_1(\tau,(1-\o_i)(z+\a\tau+\b))}{\vt_1(\tau,\o_i(z+\a\tau+\b))}
=&\cr
\apar
&\hskip 1.7cm (-1)^{\a+\b}e^{-2\pi i\ha(1-2\o_i)\a^2\tau}\(\inv{\o_i}-1\)\,,
 \quad
\hbox{if $i\in S_{\a\b}$}\,,&\?b\cr}$$}
\vskip 0.01mm
{\noindent where}
$S_{\a\b}=\{1\le i\le N\mid \o_i\a\in\bZ\ \hbox{and}\ \o_i\b\in\bZ\}$.
So the Witten index is given by
\eqn\LGOindex{
Z_{\hbox{orb}}(\tau,0)=\inv{h}\sum_{\a,\b=0}^{h-1}(-1)^N(-1)^{(D+N)(\a+\b+\a\b)}
\prod_{i\in S_{\a\b}}\(1-\inv{\o_i}\)\,.}
If $\hat c$ is an integer we can take $D=\hat c$ as mentioned before, and hence
we recover  Vafa's formula \rVafa\ for the Euler characteristic of the target
manifold up to an irrelevant factor $(-1)^N$.

Starting  with the expression of $Z(\tau,z)$ in terms of $\vt_1$
(see \LGgenus\ and \LGgenuspr)
 it  is also straightforward to evaluate the
limit
\eqn\LGOvc{\eqalign{
\lim_{\tau\->i\infty}\tsector\b\a(\tau,z)
&=(-1)^N(-1)^{(D+N)(\a+\b+\a\b)}\exp\Bigl (-2\pi iz\sum_{\o_i\a\not\in\bZ}
(\!(\o_i\a)\!)\Bigr )\cr
\apar
&\hskip 2cm\times\prod_{\o_i\a\in\bZ}
\frac{\sin\pi\{ (\o_i-1)z+\o_i\b\}}{\sin\pi\{ \o_i z+\o_i\b\}}\,,\cr}}
where $(\!(x)\!)=x-[x]-\ha$ and we can easily check the consistency between
\LGOindex\ and \LGOvc.
It is also instructive to compare \LGOvc\ with the Poincar{\' e} polynomial for
 the twisted sector \rVafa, which is left to the reader.

\newsec{Elliptic genus for manifolds with $SU(N)$ holonomy}

In this section we take $\hat c=$ integer and consider the case in which
 the charge integrality in the Neveu-Schwarz sector is already imposed.
The $N=2$ theory will then describe a complex manifold with $SU(\hat c)$
holonomy and clearly its elliptic genus must satisfy \modular\ and \period\
with $h=1$.
Once we obtain the expression of the elliptic  genus with this property
we can compute the elliptic extension of the
$\hat A$ (Dirac genus), $\sigma$ (Hirzebruch signature) and $\chi$ (Euler
characteristic) genera \rEOTY\ defined by

\eqn\specialgenus{
\eqalign{
&\hat A=q^{\hat c/8}Z(\tau,(\tau+1)/2)\,,\cr
\apar
&\sigma=Z(\tau,1/2)\,,\cr
\apar
&\chi=Z(\tau,0)\,.\cr}}

As we have seen in Sect.4, there is a systematic way to construct
the elliptic genera satisfying \modular\ and \period\ with $h=1$.
In particular, considering the elliptic genera of Landau-Ginzburg orbifolds
we obtain milliards of
concrete, but in general complicated, expressions  in terms of the
Jacobi theta function $\vt_1$
\foot{Although these expressions are of theoretical interest, they may not be
appropriate for the practical purpose of computing  expansions in $q$ and $y$
since they involve double summations over $\a$ and $\b$.}.
Instead in the subsequent Sect.5.1--5.3 we follow the methods in \rEOTY\
 to  calculate the elliptic genera for $\hat c=1,2$
and $3$ getting simple expressions
which reflects the fact that the elliptic genus is a topological invariant.
The expressions so obtained and the ones in terms of Landau-Ginzburg orbifolds
share the same modular and double quasi-periodicity properties
and hence they must coincide if the $\tau\->i\infty$ behaviors match\foot{
We have checked this by computer in several cases.}.
While  at present we do not have  similarly simple expressions for
$\hat c >3$, it is possible to
deduce general properties of the elliptic genus with $\hat c= \hbox{integer}$
and  $h=1$, and to see the results in Sect.5.1--5.3 in this light.
This will be done in Sect.5.4.

Finally, in Sect.5.5 we give a sigma model expression of the elliptic genus
and examine its properties.

\subsec{$\hat c =1$ Models}

For $\hat c =1$ the corresponding one-dimensional manifold is the complex
torus. The Landau-Ginzburg models for $\hat c =1$ are classified as the
simply elliptic singularities with the superpotentials
\eqn\?{
\eqalign{
\hat E_6 ~: & ~~~~W=X_1^3 +X_2^3 +X_3^3 + s X_1X_2X_3\,,   \cr
\apar
\hat E_7 ~: & ~~~~W=X_1^4 +X_2^4 +X_3^2 + s X_1^2 X_2^2\,, \cr
\apar
\hat E_8 ~: & ~~~~W=X_1^6 +X_2^3 +X_3^2 + s X_1^4 X_2 \,, \cr}}
where $s$ is the marginal coupling constant. Since the elliptic genus is
independent of $s$ for it is a topological invariant the elliptic genus may
be evaluated from the tensoring models.
To the $\hat E_6, ~\hat E_7$ and $\hat E_8$ models there correspond
the tensoring models $1^3, ~2^2$ and $1 \cdot 4$, respectively. We then
calculate the elliptic genus based on the method developed in \rEOTY, and
find that the elliptic genus vanishes identically. This agrees with
 the fact that the Hodge numbers for the complex torus are
$h^{0,0}=h^{0,1}=h^{1,0}=h^{1,1}=1$, and hence the Euler characteristic
vanishes.

\subsec{$\hat c =2$ Models}

It is well-known that the list of two-dimensional complex manifolds
with $SU(2)$ holonomy is exhausted by the complex two-tori and the $K3$
surface.
For the tori the elliptic genus again vanishes, and thus the $K3$ surface is
the only  non-trivial example. Fortunately the elliptic genus for the $K3$
surface has already been calculated in \rEOTY. What is left to us is to
recover the $z$-dependence in their expression. After some algebra we find
\eqn\kgenus{
Z(\tau, z)=24 \Big( {\vartheta_3(\tau, z) \over \vartheta_3} \Big)^2
       -2  {\vartheta_4^4-\vartheta_2^4 \over \eta^4}
       \Big( {\vartheta_1(\tau ,z) \over \eta} \Big)^2\,,}
where $\eta=\eta(\tau)$ and
$\vartheta_a=\vartheta_a(\tau, z=0)$ are theta
constants for $a=2,3$ and 4. One can
easily check that \kgenus\ satisfies \modular\ and
\period\ with $h=1$ .

Substituting \kgenus\ into \specialgenus\ and making $q$-expansions
we obtain
\foot{In eq.(3.6) of \rEOTY\ the coefficient 34 should read 32.}
\eqn\?{
\eqalign{
\hat A&=2 q^{-1/4}
(-1+20 q^{1/2} +62 q +216 q^{3/2} + 641 q^2 +1636 q^{5/2} + \cdots)\,,\cr
\sigma&= 16(1 + 32 q + 256 q^2 + 1408 q^3+ 6144 q^4 + 22976 q^5 +\cdots)\,, \cr
\chi&=24\,. \cr}}
Some coefficients of the double expansion of \kgenus\ with respect to
$q$ and $y=e^{2\pi iz}$ are obtained as
\eqnn\?
$$
\vbox{\halign{\hfil#\hfil&\quad\hfil#\hfil&\quad\hfil#\hfil&\quad\hfil#\hfil&\quad\hfil#\hfil&\quad\hfil#\hfil&\quad\hfil#\hfil&\quad\hfil#\hfil\cr
&$\ y^{-3}$&$\ y^{-2}$&$\ y^{-1}$&$y^{0}$&$y^{1}$&$y^{2}$&$y^{3}$ \cr
$q^{0}$&&&2&20&2 \cr
$q^{1}$&&20&-128&216&-128&20& \cr
$q^{2}$&2&216&-1026&1616&-1026&216&2 \cr}}\eqno\?
$$

\subsec{$\hat c =3$ Models}

In $\hat c =3$ $N=2$ theories with the charge integrality  imposed
there exist  extra spin 3/2 (=$\hat c /2$), $U(1)$ charge $\pm 3$, chiral
currents which together with the $N=2$ $U(1)$ current generate the $c=1$
$N=2$ algebra \rO, \rEOTY. Consequently the basic conformal blocks of
modular invariant, which are called flow-invariant orbits in \rEOTY,
are decomposed in terms
of the $c =1$ $N=2$ characters. It is crucial to note that these $c =1$ $N=2$
characters represent the $U(1)$ decoupling pieces of $\hat c =3$ $N=2$
theory. Based on this fact the flow orbits have been classified according to
the symmetry property under charge conjugation \rEOTY. This result is
sufficient for us to evaluate the elliptic genus. We find
\eqn\cygenus{
Z(\tau, z)=(h^{1,1}-h^{2,1})
  \[ f_{{1 \over 2},3}(\tau, z)+f_{{1 \over 2},3}(\tau, -z) \] \,,}
where $h^{1,1}$ and $h^{2,1}$ are Hodge numbers of a Calabi-Yau threefold
${\cal M}$ and $f_{{1 \over 2},3}$ was  defined in \defoff.
It is an immediate exercise to confirm that the properties
\modular\ and \period\ with $h=1$ are respected by \cygenus. If we let $z=0$ in
\cygenus\ we get the Euler characteristic of ${\cal M}$
\eqn\?{
Z(\tau, 0)= \op{Tr}(-1)^F = 2(h^{1,1}-h^{2,1})=\chi ({\cal M})\,,}
thanks to the Euler pentagonal identity.

\subsec{General Models}

We now wish to point out that, as suggested by the $U(1)$ decoupling argument,
the elliptic genus in the case where
$\hat c$=integer and $h=1$ can be decomposed as
\eqn\general{
Z(\tau,z) =
\sum_{m=-\hat c/2+1}^{\hat c/2} h_m(\tau) f_{m,\hat c}(\tau, z) \,,}
where we have introduced the coefficient functions $h_m(\tau)$ and the $U(1)$
theta functions $f_{m,\hat c}(\tau, z)$ whose properties are
discussed below.

First let us examine $f_{m,\hat c}$. It is defined for $\hat c$ integral
 and for $m\in \hat c/2+\bZ$ by
\eqn\newbasis{
f_{m,\hat c}(\tau, z)= {1 \over \eta (\tau )}
 \sum_{n\in \bZ} (-1)^{\hat c n}
q^{(\hat c /2) ( n+ m/\hat c) ^2}
 y^{ \hat c ( n+ m/\hat c) }\, .}
We see immediately that
\eqn\fsym{
\eqalign{
f_{m+\hat c a,\hat c}(\tau, z) &=(-1)^{\hat c a} f_{m,\hat c}(\tau, z),~~~~
a \in {\bf Z}\,,    \cr
f_{m,\hat c}(\tau, -z) &= f_{-m,\hat c}(\tau, z)\,.  \cr}}
Thus for $m= \hat c/2$
\eqn\ffsym{
f_{\hat c/2,\hat c}(\tau, -z)=(-1)^{\hat c} f_{\hat c/2,\hat c}(\tau, z)\,.}
Notice that for $\hat c$ even
\eqn\?{
f_{m,\hat c}(\tau, z)= {1 \over \eta (\tau )}
\theta_{m, \hat c/2}(\tau, 2z)\,,}
where $\theta_{m,k}(\tau, z)$ are the level-$k$ $SU(2)$ theta functions (A.1).

The functions \newbasis\ are doubly quasi-periodic with respect to $z$
\eqn\?{
f_{m,\hat c}(\tau, z+\lambda \tau +\mu)
=(-1)^{\hat c (\lambda +\mu)}
 e^{-2\pi i (\hat c/2) (\lambda^2 \tau +2\lambda z)}
f_{m,\hat c}(\tau, z)\,,}
where $\lambda, \mu \in {\bf Z}$. Notice that this is exactly the same as
\period\ with $h=1$. The modular property of \newbasis\ is deduced
as follows: under
$\tau \rightarrow \tau +1$
\eqn\?{
f_{m,\hat c}(\tau +1,z)
=e^{2\pi i (m^2/(2 \hat c) -1/24 )}
f_{m,\hat c}(\tau, z)\,,}
while under $\tau \rightarrow -1/\tau$
\eqn\?{
f_{m,\hat c}(-1/\tau, z/\tau)
=e^{2\pi i (\hat c/2) (z^2/\tau)}
\sum_{m'=-\hat c/2+1}^{\hat c/2}
B_{mm'}^{(\hat c/2)} f_{m',\hat c}(\tau, z) \,,}
where
\eqn\?{
B_{mm'}^{(\hat c/2)}= {1 \over \sqrt{\hat c}}
\exp \Big[-2\pi i {mm' \over \hat c} \Big]\,.}

We next turn to the functions $h_m(\tau)$. First,
in view of \chargesym\ and \fsym, it follows that
\eqn\hsym{
h_m(\tau)=h_{-m}(\tau)\,,}
for $|m| \leq \hat c/2-1$. In addition
$h_{\hat c/2}(\tau)=0$ for $\hat c$ odd because of \ffsym. The modular
transformation \modular\  of the elliptic genus is realized if $h_m(\tau)$
obey
\eqn\hmodular{
\eqalign{
h_{m}(\tau +1)
&=e^{-2\pi i (m^2/(2 \hat c) -1/24)}
h_{m}(\tau)\,,  \cr
h_{m}(-1/\tau)
&=\sum_{m'=-\hat c/2+1}^{\hat c/2}
  B_{mm'}^{(\hat c/2)*} h_{m'}(\tau) \,. \cr}}
Notice that the property $h_{\hat c/2}(\tau)=0$ for $\hat c$ odd is
preserved under modular transformations.
Define
\eqn\hbound{
\eqalign{
\Delta_m &= {1 \over 24}-{m^2 \over 2 \hat c} ~,  \cr
\chi_{p} &= \sum_{q=0}^{\hat c} (-1)^{p+q} h^{p,q},
                   ~~~~p=0,1,2,\cdots ,\hat c\,,  \cr}}
where $h^{p,q}$ are the Hodge numbers of the corresponding manifold
${\cal M}$ with $SU(\hat c)$ holonomy.
Inspecting the
$\tau \rightarrow i\infty$
behavior of $f_{m,\hat c}(\tau, z)$ and the Ramond vacuum
charge configuration
we find that the $q$-expansion of
$h_m(\tau)$ takes the form
\eqn\hlimit{
h_m(\tau)=\chi_{m+\hat c/2}~q^{\Delta_m} + \cdots \,,}
where
the ellipsis contains the terms
with powers $\Delta_m+\bZ_{>0}$
\foot{Because of the property that $\Delta_m -1/24 < 0$ for $m\neq 0$
our elliptic genus is not quite a Jacobi form which is studied in \rEZ.}.
Then we obtain
\eqn\?{
\lim_{\tau \rightarrow i\infty} Z(\tau, z)
=\sum_{p=1}^{\hat c-1}\chi_{p}~ y^{-\hat c/2+p}
 +\chi_{\hat c}~y^{\hat c /2} +\chi_{0}~y^{-\hat c/2}\,,}
where we have used the Poincar\'e duality $\chi_{\hat c}=\chi_0$ and, for
$\hat c$ odd, $\chi_{\hat c}=\chi_0=0$. The Witten index may be evaluated as
\eqn\?{
\op{Tr}(-1)^F = \lim_{\tau \rightarrow i\infty} Z(\tau, 0)
                  = \sum_{p=0}^{\hat c}\chi_{p}
                 =\sum_{p,q=0}^{\hat c}(-1)^{p+q} h^{p,q}
                  =\chi({\cal M}) \,.}
Thus \general\ has the desired property of the
elliptic genus if $h_m(\tau)$ are subject to \hsym-\hlimit.

Let us examine \general\ for $\hat c=1,~2$ and 3. First, for $\hat c=1$,
we have only $m=1/2$, and hence the elliptic genus vanishes in agreement
with our previous observation in Sect.5.1.
Next we notice for $\hat c=2$ that $f_{m,1}(\tau, z)$ reduce to the level-1
$SU(2)$ Kac-Moody characters for the highest weight representations with
isospin 0 ($m=0$) and 1/2 ($m=1$). The $SU(2)$ characters appear
since, under the charge integrality condition, the $\hat c=2$ $N=2$
theories exhibit enhanced symmetry of the $N=4$ superconformal
algebra which contains the $SU(2)$ algebra as its sub-algebra. The value
$\hat c=2$ is the lowest allowed value corresponding to the level-1 $SU(2)$
algebra in the $N=4$ unitary representations \rET.
We can read off  $h_0(\tau)$ and $h_1(\tau)$ from \kgenus. The result is
\eqn\?{
\eqalign{
h_0(\tau) &=24 {\eta \, \theta_{0,1} \over \vartheta_3^2}
-2{\vartheta_4^4-\vartheta_2^4 \over \eta^4}{\theta_{1,1} \over \eta}  \cr
\apar
&=q^{1\over 24} \big(20+196q+1380q^2+6200q^3+23400q^4+76220q^5+\cdots \big)
 \,,\cr}}
%
and
\eqn\?{
\eqalign{
h_1(\tau) &=24 {\eta \, \theta_{1,1} \over \vartheta_3^2}
+2{\vartheta_4^4-\vartheta_2^4 \over \eta^4}{\theta_{0,1} \over \eta} \cr
\apar
&=q^{-{5\over 24}} \big(2-130q-900q^2-4350q^3-17020q^4-57344q^5+\cdots \big)
 \,, \cr}}
%
where $\theta_{m,1}=\theta_{m,1}(\tau,0)$. One checks easily \hmodular\
as well as \hlimit\ with \hbound\ since
$h^{1,1}=20,~h^{2,0}=h^{2,2}=1,~h^{1,0}=h^{1,2}=0$ for the K3 surface.
Finally, turning to the $\hat c=3$ case, we find that the $3\times 3$ matrix
$B_{mm'}^{(3/2)*}$ in \hmodular\ has an eigenvector
${}^t(1,~1,~0)$ with the eigenvalue unity. Thus we take
\eqn\?{
(h_{-1/2}(\tau),~h_{1/2}(\tau),~h_{3/2}(\tau))
=(h^{1,1}-h^{2,1})(1,~1,~0)\,,}
which obviously satisfy \hsym-\hlimit. The result indeed agrees with
\cygenus\ in Sect.5.3.

\subsec{Sigma model expression}

Let $\CM$ be a  ${\hat c}$ dimensional K{\" ahler}  manifold
and denote the total Chern class of  $\CM$ by
\eqn\?{
c(\xi)=\sum_{j=0}^{\hat c}c_j(\xi)=\prod_{j=1}^{\hat c}(1+\xi_j)\,,}
where $\xi_j$ are the skew eigenvalues of the curvature two form
$\frac{i}{2\pi}\CR_{a \bar b}$ \rEGH.
We propose the following formula\foot{See
\rHirzebruch\ for a similar expression.}
as a sigma model expression of  the elliptic genus for $\CM$:
\eqn\SMgenus{
Z(\tau,z)=\int_\CM\,G(\tau,z,\xi)\,,}
where
\eqn\SMintegrand{
G(\tau,z,\xi)=\prod_{j=1}^{\hat c}\,
\frac{\vt_1(\tau,\xi_j+z)}{\vt_1(\tau,\xi_j)}\,\xi_j\,.}
Note that in \SMgenus,
only homogeneous terms of degree $\hat c$ in $\xi_j$ survive when
integrating over $\CM$.
The modular properties and double quasi-periodicity  of $G$ are
given by
\eqn\trofG{
\eqalign{
&G\(\frac{a\tau+b}{c\tau+d},\frac{z}{c\tau+d},\frac{\xi}{c\tau+d}\)\cr
\apar
&\hskip 2cm=e^{ 2\pi i c_1(\xi) cz/(c\tau+d)}\,(c\tau+d)^{-\hat c}\,
e^{2\pi i (\hat c /2) c z^2 /( c\tau +d)} G(\tau, z, \xi)\,,\cr
\apar
&G(\tau, z+\lambda \tau +\mu,\xi)=e^{-2 \pi i c_1(\xi)\la}
(-1)^{\hat c (\lambda +\mu)}
 e^{-2\pi i (\hat c/ 2) (\lambda^2 \tau +2\lambda z) }
G(\tau, z, \xi)\,,\cr
\apar
&\hskip 3cm\pmatrix{a&b\cr c&d\cr}\in SL(2,\bZ)\,,\quad
\lambda, \mu \in {\bf Z}\,,\cr}}
where $c_1(\xi)=\sum_{j=1}^{\hat c} \xi_j$ is the first Chern class.

Thus we observe that whenever the first Chern class does not vanish
the elliptic genus
\SMgenus\ suffers from the global anomaly \rLW\
while if $\CM$ has $SU(\hat c)$ holonomy, then,
 by virtue of $c_1(\xi)=0$, \SMgenus\  now satisfies
\modular\ and \period\ with $h=1$.
This result is in accordance  with the long supported anticipation that
a sigma model, whose target space is a K{\" a}hler manifold with vanishing
first Chern class, allows a description as $N=2$ SCFT.
By substituting \SMgenus\ into \specialgenus\ we find again by use of
$c_1(\xi)=0$ that
\eqn\SMspecialgenus{
\eqalign{
\hat A&=\,\int_\CM\,\prod_{j=1}^{\hat c}
\frac{\vt_3(\tau,\xi_j)}{\vt_1(\tau,\xi_j)}\,\xi_j\,
=q^{-\hat c/8}\,\int_\CM\,\prod_{j=1}^{\hat c}\frac{\xi_j/2}{\sin\pi\xi_j}+
\cdots\,,\cr
\apar
\sigma&=\,\int_\CM\,\prod_{j=1}^{\hat c}\,
\frac{\vt_2(\tau,\xi_j)}{\vt_1(\tau,\xi_j)}\,\xi_j=
\,\int_\CM\,\prod_{j=1}^{\hat c}\frac{\xi_j}{\tan\pi\xi_j}+\cdots\,,\cr
\apar
\chi&=\,\int_\CM\,\prod_{j=1}^{\hat c}\,\xi_j\,,\cr}}
In  \SMspecialgenus\ we discern the classical expressions in the first terms of
$q$-expansions.  The results for the $K3$ surface in Sect.5.2 agree
 with \SMgenus.

Finally  we wish to point out  that there exists another expression of
\SMgenus\
in terms of  Dirac determinants on the two-dimensional torus. We have\foot{
See the second reference by
Schellekens and Warner \rSW\ for more details.}
\eqn\Detgenus{
\eqalign{
G(\tau,z,\xi)&=y^{-\hat c/2}\prod_{j=1}^{\hat c}
\frac{\Det_z'(\del+\xi_j)}{\Det'(\del+\xi_j)}
\frac{\Det(\bar\del+\xi_j)}{\Det'(\bar\del+\xi_j)}\cr
\apar
&=y^{-\hat c/2}
\prod_{j=1}^{\hat c}\frac{\Det_z'(\del+\xi_j)\Det(\bar\del+\xi_j)}
{\Det'(\del\bar\del+\xi_j\del+\xi_j\bar\del+\xi_j^2)}\,,\cr}}
where all the determinants are evaluated with periodic-periodic conditions
except for $\Det_z'$ for which we assume periodicity in the space direction but
a twisted boundary condition by $e^{2\pi iz}$ in the time direction.
In \Detgenus\
the primes over determinants represent deletion of the eigenvalues
corresponding to the zero eigenvalues of $\del$ and $\bar \del$ operators.
This representation should be understandable via  a  path integral approach.

\ack{We thank A.~Fujitsu for discussion on free field realizations.}

\appendix{A}{Properties of theta functions}

For a positive integer $k$   theta functions of level $k$ are defined
by
\eqn\?{
\eqalign{
\t_{m,k}(\tau,z)&=
\sum_{n\in\bZ}q^{k(n+\frac{m}{2k})^2}y^{k(n+\frac{m}{2k})}\,.\cr
\apar
m&=-k+1,\ldots,k\,,\cr}}
where $q=e^{2\pi i\tau}$ and $y=e^{2\pi i z}$.
They transform under modular transformations as
\eqna\?{
$$\eqalignno{
&\t_{m,k}(\tau+1,z)=e^{2\pi i\frac{m^2}{4k}}\t_{m,k}(\tau,z)\,,&\? a\cr
\apar
&\t_{m,k}(-1/\tau,z/\tau)=e^{2\pi i(z^2/4\tau)}(-i\tau)^{1/2}\sum_{m'=-k+1}^k
B_{mm'}^{(k)}\t_{m',k}(\tau,z)\,,&\? b\cr
\apar
&\t_{m,k}(\tau,-z)=\t_{-m,k}(\tau,z)\, ,&\? c\cr}$$
}
where
\eqn\?{B_{mm'}^{(k)}=\inv{\sqrt{2k}}e^{-\pi i mm'/k}\, .}
The following is a slight extension of the usual multiplication formula,
\eqn\multformula{\eqalign{
\t_{m,k}(\tau,z)&\t_{m',k'}(\tau,z')=\cr
\apar
& \sum\limits_{j\in\bZ/(k+k')\bZ}
\t_{mk'-m'k+2kk'j,kk'(k+k')}(\tau,u)
\t_{m+m'+2kj,k+k'}(\tau,v)\,,\cr}}
where $u=(z-z')/(k+k')$ and $v=(kz+k'z')/(k+k')$.

The Jacobi theta function $\vt_1(\tau,z)$ is defined by
\eqn\?{\eqalign{
\vt_1(\tau,z)&= -i(\t_{1,2}-\t_{-1,2})(\tau,z)=
i \sum_{n\in\bZ}(-1)^n q^{\ha(n-\ha)^2}y^{n-\ha}\cr
\apar
&=iq^{\inv{8}}y^{-\ha}\prod_{n=1}^\infty(1-q^n)(1-q^{n-1}y)(1-q^ny^{-1})\,,\cr}}
and satisfies the modular transformation laws
\eqna\?{
$$\eqalignno{
&\vt_1(\tau+1,z)=e^{2\pi i(1/8)}\vt_1(\tau,z)\,,&\? a\cr
\apar
&\vt_1(-1/\tau,z/\tau)=(-i\tau)^{1/2} e^{2\pi i(1/2)(z^2/\tau)}
\vt_1(\tau,z)\,,&\? b\cr
\apar
&\vt_1(\tau,-z)=-\vt_1(\tau,z)\,,&\? c\cr}$$}
as well as the double quasi-periodicity
\eqn\?{
\vt_1(\tau,z+\a\tau+\b)=(-1)^{\a+\b}e^{-2\pi i(1/2)(\a^2+2\a z)}\vt_1(\tau,z),
\quad\a,\b\in\bZ\, .}
As a function of $z$, $\vt_1(\tau,z)$ has no poles but has simple zeros:
\eqn\?{
\vt_1(\tau,\a\tau+\b)=0,\quad\a,\b\in\bZ\, .}
The remaining Jacobi theta functions are defined by
\eqn\?{
\eqalign{
\vt_2(\tau,z)&=\sum_{n \in {\bf Z}}
          q^{{1 \over 2}(n-\ha)^2} y^{n-{1 \over 2}}\,, \cr
\apar
\vt_3(\tau,z)&=\sum_{n \in {\bf Z}}
          q^{{1 \over 2} n^2} y^{n}\,, \cr
\apar
\vt_4(\tau,z)&=\sum_{n \in {\bf Z}}
          (-1)^n q^{{1 \over 2} n^2} y^{n}\,. \cr}}

\appendix{B}{Basic properties of $I_m^l$ }

Here we summarize basic properties of $I_m^l$:
\eqna\propofI{
$$\eqalignno{
&I_m^l(\tau+1,z)=e^{2\pi i (h_m^l-(3\hat c)/24)}I_m^l(\tau,z)\,,
&\propofI a\cr
\apar
&I_m^l(-1/\tau,z/\tau)= (-i) e^{2\pi i (\hat c/2) (z^2/\tau)}
\sum_{l'=0}^k\sum_{m'=-k-1}^{k+2}A^{(k)}_{ll'}
B_{mm'}^{(k+2)*}I^{l'}_{m'}(\tau,z)\,, &\propofI b\cr
\apar
&I^l_m(\tau,-z)=-I^l_{-m}(\tau,z)=I^{k-l}_{k+2-m}(\tau,z)\,,&\propofI c\cr
\apar
&I^l_m(\tau,z)=I^l_{m+2(k+2)\bZ}(\tau,z)\,,&\propofI d\cr
\apar
&I^l_m(\tau,z+\a\tau+\b)=(-1)^{\a+\b}e^{2\pi i\frac{m}{k+2}\b}
e^{-2\pi i(\hat c/2)(\a^2\tau+2\a z)}I^l_{m-2\a}(\tau,z)\,,\quad
\a,\b\in\bZ\,,&\cr
&&\propofI e\cr}$$}
where $h_m^l\equiv\frac{l(l+2)-m^2}{4(k+2)}+\inv{8}\pmod{\bZ}$ in \propofI{a}.

\listrefs

\bye